\documentclass[runningheads]{llncs}
\usepackage[T1]{fontenc}
\usepackage{xcolor}
\usepackage{graphicx}
\usepackage{amsmath}
\usepackage{algorithm}
\usepackage[noend]{algorithmic}
\usepackage{multirow}
\usepackage{amssymb}
\usepackage{booktabs} 
\usepackage{amssymb} 
\usepackage{hyperref}
\usepackage{marvosym}
\begin{document}
%
\title{Phy-Diff: Physics-guided Hourglass Diffusion Model for Diffusion MRI Synthesis\thanks{J. Zhang and R. Yan contribute equally in this work.}}

\titlerunning{Phy-Diff for dMRI Synthesis}
%
\author{Juanhua Zhang\inst{1} \and
Ruodan Yan\inst{2} \and
Alessandro Perelli\inst{1} \and
Xi Chen\inst{3}\and
Chao Li\inst{1,2,4,5}\textsuperscript{\Letter}}

\authorrunning{J. Zhang et al.}
%
\institute{
School of Science and Engineering, University of Dundee
\and
Department of Applied Mathematics and Theoretical Physics, University of Cambridge
\and
Department of Computer Science, University of Bath
\and
Department of Clinical Neurosciences, University of Cambridge
\and 
School of Medicine, University of Dundee
\\
\email{cl647@cam.ac.uk}
}
\maketitle              
\begin{abstract}
Diffusion MRI (dMRI) is an important neuroimaging technique with high acquisition costs. Deep learning approaches have been used to enhance dMRI and predict diffusion biomarkers through undersampled dMRI. To generate more comprehensive raw dMRI, generative adversarial network based methods are proposed to include b-values and b-vectors as conditions, but they are limited by unstable training and less desirable diversity. The emerging diffusion model (DM) promises to improve generative performance. However, it remains challenging to include essential information in conditioning DM for more relevant generation, i.e., the physical principles of dMRI and white matter tract structures. In this study, we propose a physics-guided diffusion model to generate high-quality dMRI. Our model introduces the physical principles of dMRI in the noise evolution in the diffusion process and introduces a query-based conditional mapping within the diffusion model. In addition, to enhance the anatomical fine details of the generation, we introduce the XTRACT atlas as a prior of white matter tracts by adopting an adapter technique. Our experiment results show that our method outperforms other state-of-the-art methods and has the potential to advance dMRI enhancement.

\end{abstract}

\section{Introduction}
Diffusion MRI (dMRI) is an essential neuroimaging modality due to its sensitivity in probing white matter fibers and characterizing neurological diseases \cite{le_bihan_diffusion_2012,WeiYiranBrainTumor}. However, acquiring high-quality dMRI for sufficient fidelity remains a challenge due to expensive acquisition costs. For example, the neurite orientation dispersion and density imaging (NODDI) sequence requires 90 diffusion gradients, which is time-consuming and infeasible for patients with severe conditions \cite{zhang_noddi_2012}. There is an increasing need to develop effective algorithms for dMRI enhancement to reduce acquisition costs. 

Deep learning has achieved encouraging performance in medical image synthesis. Based on conventional MRI, various frameworks have been proposed with success to synthesize multimodal MRI \cite{jiang_cola-diff_2023}, super-resolve images \cite{GANReDL}, and generate brain connectivities \cite{BrainNetGAN,wei2023multi}, supporting the feasibility of deep learning dMRI enhancement. Several studies have proposed frameworks to estimate dMRI parametric maps from undersampled high-resolution dMRI. For instance, a CNN-based model was proposed using a few dMRI directions to generate diffusion kurtosis imaging (DKI) and NODDI maps \cite{golkov_q-space_2016}. Further, other methods were proposed based on Graph Transformer and atlas-aided U-Net to synthesize dMRI biomarkers \cite{chen_hybrid_2022,karimi_atlas-powered_2022}. However, these methods use sparsely sampled dMRI to generate specific parametric maps without leveraging b-values and b-vectors from raw dMRI for more comprehensive image generation. Therefore, these methods require retraining when applied to generating other parametric maps.

To mitigate this challenge, generative adversarial network (GAN) based methods are proposed using b-values and b-vectors as conditions, promising to generate comprehensive raw dMRI. The generated images are used to calculate downstream parametric maps \cite{ren_q-space_2021}. Despite the success, GAN-based architectures are often limited by their inability to balance fine details and visual quality \cite{bau_seeing_2019}, which could be suboptimal for clinical diagnosis. In addition, GANs capture less diversity in large-volume generation \cite{dhariwal_diffusion_2021}, which cannot meet real-world clinical needs to address patient individuality. In addition, GANs often demonstrate unstable training, challenged by gradient vanishing and mode collapse \cite{optimization_gan}.

As a type of likelihood-based model, diffusion models (DM) promise to achieve better generation performance while addressing the limitations of GANs \cite{dhariwal_diffusion_2021}. Mounting research has shown that DMs are successful in generating high-quality, complex, and diverse images \cite{ddpm,DisC-Diff,improvedDDPM}, with properties of fixed training targets and easy scalability \cite{no_modecollapse_ddpm}. Despite the advantages, there remains an unmet need to enhance dMRI effectively due to several challenges: 1) dMRI is distinct from conventional MRI in underlying physics principles. Therefore, generating dMRI may require additional physics-informed modeling beyond image geometry. However, typical DMs only add or remove random Gaussian noise in the diffusion process; 2) It remains a significant challenge for DMs to preserve the anatomical structures that are essential to the efficacy of dMRI in assessing tissue microstructures and characterizing diseases. This is particularly important for the white matter tract regions that dMRI mainly reflects; 3) It is desirable to improve DMs to generate dMRI in a controllable manner with multiple conditions required, which is essential for restoring anatomical details in dMRI \cite{dhariwal_diffusion_2021}. 

We propose a DM-based generative model tailored for dMRI, promoting image translation from the baseline images without gradient applied, i.e., $b_0$ images. Specifically,  we introduce a physics-guided noise evolution and a query-based condition mapping. To better preserve the structural integrity of the white matter tracts in the synthesized dMRI, we introduce local conditions of the XTRACT atlas, with 42 anatomical tracts constructed from more than 1,000 subjects \cite{warrington_xtract_2020}. Our main contributions include:

\begin{itemize}
\item We develop a physics-guided diffusion model to synthesize high-quality dMRI. As far as we know, this is the first attempt to generate raw dMRI using the conditional diffusion model. 
\item We propose a physics-guided noise evolution process in the forward diffusion to reflect the noise characteristics inherent in dMRI, which could improve on traditional DMs adding Gaussian noise. 
\item We design a query-based conditional mapping in the reverse diffusion process, incorporating q-space sampling to enable continuous diffusion gradient mapping, allowing for better directional dMRI synthesis at a given gradient.

\item We introduce XTRACT atlas \cite{warrington_xtract_2020} as anatomical prior in conditioning DM for preserving the fine details of tract. Inspired by the T2I-Adapter \cite{mou_t2i-adapter_2023}, we improve the model adaptability for more controllable image generation while allowing to incorporate multiple conditions.
\end{itemize}

According to experimental results, our model achieves state-of-the-art performance and could serve as a promising tool for dMRI enhancement.

\begin{figure}
\centering
\includegraphics[width=0.925\textwidth]{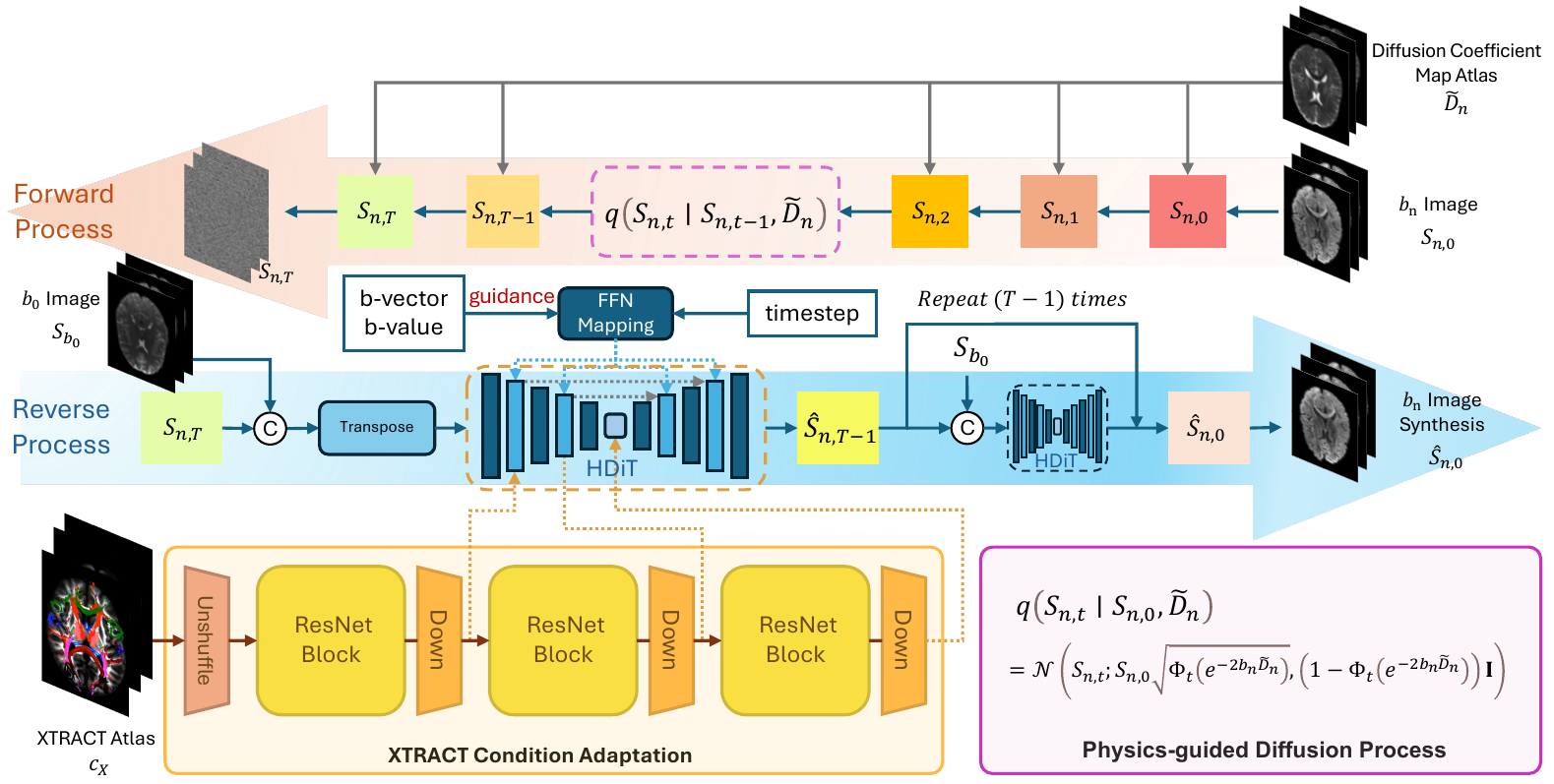}
\caption{\textbf{Model architecture.}\; The physics information ADC atlas $\tilde{\mathcal{D}}_n$ interacts with the original image $b_n$ during the forward process by the function $q\left(S_{n,t}|S_{n,0},\tilde{\mathcal{D}}_n\right)$. After timesteps $T$, the image turns into the $S_{n,T}$. During the reverse process, HDiT predicts the added noise in each timestep in the pixel space, while using the b-vector, b-value as conditions via query-based conditional mapping. After training the noise prediction network, the XTRACT adapter is used to incorporate tractography structure $c_X$. By sampling from the learned pattern, the synthesized image $\hat{S_{n,0}}$ is obtained.}
\label{fig.arch}
\end{figure}

\section{Methodology}
\subsubsection{Overview.}
Fig.~\ref{fig.arch} illustrates the model architecture. We propose a physics-guided diffusion process based on the framework of Denoising Diffusion Probabilistic Models (DDPM) \cite{ddpm}, including a physics-guided noise evolution process and a query-based conditional mapping mechanism in the inference process. In the noise evolution process, Phy-Diff incorporates Apparent Diffusion Coefficient (ADC) maps to learn the underlying noise representation in the synthesized dMRI data ($b_n$) (see~\ref{noise.evolution}). During the reverse process, we employ a query-based conditional mapping technique that incorporates q-space sampling to guide the generation of $b_n$ under the Hourglass Diffusion Transformer (HDiT) backbone, facilitating the more precise directional dMRI synthesis (see~\ref{hdit}). Further, a XTRACT condition adaptation is designed to provide anatomical conditions that guide the structural integrity of tracts in synethsized dMRI (see~\ref{xtract.adapt}).

\subsection{Physics-guided Noise Evolution}
\label{noise.evolution}
The proposed physics-guided noise evolution process learn the underlying noise representation by leveraging an ADC atlas, where atlas refering to an average map. We first consider $x_{t}$ as the $t$-th image representation, and the approximate posterior $q\left(x_t\mid x_0\right)$, representing the \emph{diffusion process}, can be expressed as:
\begin{equation}
q\left(x_t\mid x_0\right)=\mathcal{N}\left(x_t;\sqrt{\bar{\alpha}_t}x_0,\left(1-\bar{\alpha}_t\right)\mathbf{I}\right)
\end{equation}
where, $x_t$ is the noisy image at the timestep $t$, $x_0$ is the image that is expected to be generated, and ${{\bar \alpha }_t} = \prod\nolimits_{i = 1}^t {{\alpha _i}}$, and ${{\alpha }_i}$ are hyper-parameters relevant to the variance. Thus, $x_{t}$ can be computed by:
\begin{equation}
x_t=\sqrt{\bar{\alpha}_t} x_0+\sqrt{1-\bar{\alpha}_t} \epsilon, \quad \text { with } \epsilon \sim \mathcal{N}(0, \mathbf{I})
\label{ddpm_forward}
\end{equation}
where $\epsilon$ represents the Gaussian noise schedule. By gradually increasing the proportion of noise and decreasing the proportion of content on the image, the final image will become completely noisy. The backbone network of DM gradually removes noise by predicting the noise changes at each timestep, ultimately restoring the original image.

However, the denoising mechanism of the standard DDPM may be more suitable for natural image generation, by multiplying the Gaussian noise with the uniform distributed variance, less considering the specific structure of dMRI. To address this, we incorporate the diffusion physics inherent in dMRI into the process. Specifically, the diffusion physics information is integrated into the forward phase of DDPM:
\begin{equation}
S_n=S_0 e^{-b_n\mathcal{D}_n}+R(\mathcal{D}_n)
\label{dwi_physics}
\end{equation}
where $\mathcal{D}_n$ represents diffusion coefficient, $b_n$ is an arbitrary b-value, and $S_n$ and $S_0$ denote the signal intensities corresponding to the diffusion-weighted images under b-values $b_n$ and $b_0$, respectively. In order to approach the formula of DDPM as closely as possible, we developed a residual term $R(\mathcal{D}_n)$ to effectively correct the modeling errors caused by signal errors.

We then integrate this physical process (Eq.~\ref{dwi_physics}) into the diffusion model architecture by assuming $\bar{\alpha}_t=\Phi_t\left(e^{-2b_n\mathcal{D}_n}\right)$ and $R(\mathcal{D}_n)=\sqrt{1-\Phi_t\left(e^{-2b_n\mathcal{D}_n}\right)} \epsilon$, where $\epsilon$ is the residual schedule. Specifically, $\Phi_t(\cdot)$ refers to a transformation used to map the range of the function argument, making the diffusion process more stable. The goal is to solve for $S_n$. Therefore, we relax the restriction on $\mathcal{D}_n$ and estimate that in the original equation as $\tilde{\mathcal{D}}_n$ by introducing ADC atlas, which is expressed as:
\begin{equation}
\tilde{\mathcal{D}}_n={\sum_{i=1}^{N}{\mathcal{D}^i_n}}={\frac{N\ln{S_0}-\ln{\prod_{i=1}^{N}{S^i_{n}}}}{b_n}}
\end{equation}
where $N$ denotes the number of all dMRI images $\{S^0_n,S^1_n,\ldots,S^N_n\}$ under the same b-value $b_n$, and $S_0$ refers to the dMRI image under $b=0$.

Consequently, our physics-guided noising mechanism is formulated as:
\begin{align}
S_{n,t}&=S_{n,0}\sqrt{\Phi_t\left(e^{-2b_n\tilde{\mathcal{D}}_n}\right)}+\sqrt{1-\Phi_t\left(e^{-2b_n\tilde{\mathcal{D}}_n}\right)} \epsilon
\end{align}
The reverse process is also similar. Ultimately, the simplified form of the goal we need to optimize should be:
\begin{equation}
\mathcal{L}_{\mathrm{simple}}\left(\theta\right)=\mathbb{E}_{\epsilon\sim\mathcal{N}\left(0,1\right)}\left[\left\|\epsilon-D_\theta\left(S_{n,t},t,\overrightarrow{b_n},p_s,b_n\right)\right\|_2^2\right]
\end{equation}
where, $t$ is the diffusion timestep. In addition, function $D_\theta(\cdot)$ represents the backbone neural network model for noise prediction, and $\overrightarrow{b_n}$, $p_s$ as well as $b_n$ are the b-vector, slice index and the b-value guidance, which will all be explained in detail in the following text.

\subsection{Query-based Conditional Mapping}
\label{hdit}
We introduce a query-based conditional mapping technique within the HDiT framework, to predict the noise at each timestep $t$. We employ HDiT due to its efficiency in generating high-quality images in the pixel space because of the use of Neighborhood Attention (NA) \cite{crowson2024scalable}. 

The q-space condition is defined as the b-vector $\overrightarrow{b_n}=\left(b_x,b_y,b_z\right)$ and its magnitude b-value $|\overrightarrow{b_n}|$, thus the q-space sampling, unlike integer scalar index categories, e.g., real number or vector categories, requires tailored embedding to use as condition. Here, we develop a category embedding method for query-based conditional mapping, to further capture the directional information of the physics guidance $\tilde{D}_n$. In detail, b-vectors are normalized and fed into a multi-layer perceptron (MLP) based positional embedding for stability, but overlooking the magnitude information specifying dMRI intensity, so the b-values are also introduced as guidance. Since traditional class embedding only applies to the input of integer scalar indexes as categories, ignoring the signal strength information represented by real numbers, an embedding method that indicates magnitude categories based on real numbers is proposed. The real number embedding function $\varepsilon_{real}(\cdot)$ can be briefly represented as:
\begin{equation}
\varepsilon_{real}\left(x\right)=f_{MLP}\left[\frac{x}{\left|x\right|}\ln\left(\left|x\right|+1\right)\right]
\end{equation}
where $f_{MLP}(\cdot)$ represents the MLP function. This approach could benefit numerical stability.

Subsequently, guidance embeddings, including embedded slice index, are integrated into the Transformer with $N$ feed-forward network (FFN) blocks, using GEGLU \cite{shazeer2020glu}. The process includes enhancing the image with an MLP, three NA HDiT blocks for downsampling, a self-attention Transformer, and three more NA HDiT blocks for upsampling, all interacting with the guidances \cite{crowson2024scalable}.

\subsection{XTRACT Condition Adaptation}
\label{xtract.adapt}
It is essential to preserve tract information for dMRI generation, thus we introduce the XTRACT, which includes 42 anatomically dissected tracts derived from over 1,000 human subjects. To better incorporate these tracts into the DM conditioning, we employ an adapter technique inspired by \cite{warrington_xtract_2020}, which could improve the adaptability of the model for controllable image generation while incorporating multiple conditions.

Particularly, in order to fully adapt the XTRACT atlas to DM, it is necessary to exclude meaningless contents. We note that not all the image slices have tracts on the XTRACT atlas. Therefore, we develop a module to enrich the empty positions. For those empty XTRACT, we combine $S_{b_0}$ with the sum of the non-empty tracts, which can be simply expressed as:
\begin{equation}
c_{X\left(i,j\right)}^{\prime}=\left(\sum_{k=1}^{42}c_{X\left(i,k\right)}+\xi\right)\cdot f_N\left(S_{b_0}\right)
\end{equation}
where $c_{X\left(i,j\right)}$ means the XTRACT condition in the $i$-th slice and the $j$-th tract, $c_{X\left(i,j\right)}^{\prime}$ is its replacement, $\xi$ is a correction factor, and $f_N(\cdot)$ represents the min-max normalization.

Subsequently, as illustrated in Fig.~\ref{fig.arch}, we use a pixel unshuffle with a downsampling factor of $4$ to downsample the input XTRACT atlas as conditions. Then, three classic ResNet blocks with each followed downsampling blocks are adopted. All features from each downsampling procedure are permuted to be consistent with each corresponding HDiT block for further addition.

\section{Experiments and Results}
\subsection{Datasets and Preprocessing}
We collected dMRI data from 9 randomly selected subjects from the Human Connectome Project (HCP) S1200 \cite{essen_wu-minn_2013}. These subjects were divided into training, validation and testing sets in a ratio of 7:1:1. Each subject contains images across b-value shells at 1000, 2000 and 3000 $s/mm^2$, with each shell comprising 90 diffusion directions along with $b_0$ images. Furthermore, 110 2D slices are extracted from each 3D images.

To ensure spatial consistency with XTRACT atlas, we first registered dMRI images to their corresponding T1-weighted images, followed by non-linearly transformed to the standard space by registering them to MNI-305 \cite{FONOV2011313} using the Advaned Normalization Tools (ANTs). Additionally, we normalized all dMRI images, ADC atlas and XTRACT atlas, as well as b-vectors into the range of $[-1,1]$, and  padded all input images to ensure their resolution at $256\times256$. 

\subsection{Implementation Details} 
We implement the following hyperparameters: linear noise schedule; $1000$ diffusion steps; noise variances at the range of $\beta_1=10^{-4}$ and $\beta_T=0.02$; patch size of $4\times4$; batch size of $32$; AdamW optimizer with learning rate at $5.0\times10^{-4}$, weight decay at $10^{-4}$, betas at $0.9$ and $0.95$, eps at $10^{-8}$. The HDiT are implemented with three layers with dimensions of $128$, $256$, $512$; 2 Transformer blocks in each layer. The Adapter convolution channels are the same as the HDiT dimensions, with a kernel size of $1$. The model is trained in a NVIDIA A100-PCIE-40GB, and implemented through PyTorch. Maximum epochs of 80 is set and early stopping is used.

\subsection{Comparisons with SOTA and Baselines}
We compared our model with 5 state-of-the-art DWI synthesis approaches (Table~\ref{comparison}): U-Net \cite{ronneberger2015unet}, U-Net++ \cite{zhou2018unet}, q-DL \cite{golkov_q-space_2016}, q-DL (2D) \cite{qDLCNN}, SMRI2DWI \cite{ren_q-space_2021}, LDM \cite{rombach_high-resolution_2022} and the original HDiT \cite{crowson2024scalable}. As conditioning is essential for our task, all models underwent slight modifications to incorporate basic conditioning, enabling them to utilize b-vectors and slice indexes as conditions. Specifically, they were fed into the vanilla MLP and further embedded into the model backbones.

\begin{table}[t]
  \centering
  \caption{Performance on the HCP S1200 dataset. \textbf{Boldface} marks the top models. The b-values between arbitary $b_n$, $b_n=1000\;s/mm^2$, $b_n=2000\;s/mm^2$ and $b_n=3000\;s/mm^2$ as conditions are tested.}
  \resizebox{\linewidth}{!}{
    \begin{tabular}{lllllllll}
    \toprule
    \multicolumn{1}{c}{\multirow{3}[4]{*}{Model}} & \multicolumn{2}{c}{Arbitrary $b_n$} & \multicolumn{2}{c}{$b_n=1000\;s/mm^2$} & \multicolumn{2}{c}{$b_n=2000\;s/mm^2$} & \multicolumn{2}{c}{$b_n=3000\;s/mm^2$} \\
\cmidrule{2-9} & SSIM\% & PSNR & SSIM\% & PSNR & SSIM\% & PSNR & SSIM\% & PSNR \\
    \midrule
    U-Net & 61.92±11.3 & 23.90±3.08 & 66.07±10.5 & 25.48±2.19 & 65.64±11.2 & 25.00±2.16 & 65.91±10.6 & \textbf{25.20±2.84} \\
    U-Net++ & 61.89±9.34 & 23.21±2.83 & 65.74±10.8 & 25.20±2.43 & 66.31±10.3 & 24.49±2.19 & 65.18±12.3 & 24.21±3.03 \\
    q-DL & 78.59±7.16 & 13.24±2.65 & 80.45±8.54 & 15.54±2.20 & 79.97±5.09 & 14.92±1.80 & 78.57±11.1 & 15.11±2.34 \\
    q-DL 2D & 63.41±21.25 & 21.93±3.38 & 67.88±18.6 & 25.41±2.85 & 70.63±16.66 & \textbf{25.52±2.94} & 66.10±21.7 & 24.78±2.70 \\
    SMRI2DWI (CGAN) & 32.12±8.89 & 20.71±3.15 & 33.32±7.48 & 24.53±2.38 & 34.66±6.10 & 24.02±2.56 & 34.67±7.80 & 24.06±2.78 \\
    Latent Diffusion & 70.69±8.66 & 21.31±2.11 & 69.96±9.08 & 21.00±1.47 & 68.69±7.59 & 20.89±1.51 & 69.26±8.50 & 21.31±1.66 \\
    DDPM (HDiT) & 81.92±5.41 & 22.70±2.26 & 86.78±3.94 & 25.72±1.95 & 82.71±4.92 & 23.36±2.02 & 81.19±6.51 & 22.26±3.01 \\
    \textbf{Phy-Diff (Ours)} & \textbf{84.27±6.17} & \textbf{24.46±2.12} & \textbf{87.64±4.14} & \textbf{25.83±2.20} & \textbf{83.10±5.83} & 23.68±2.55 & \textbf{81.44±5.82} & 23.59±1.54 \\
    \bottomrule
    \end{tabular}
    }
  \label{comparison}
\end{table}

\begin{figure}
    \centering
    \includegraphics[width=0.9\textwidth]{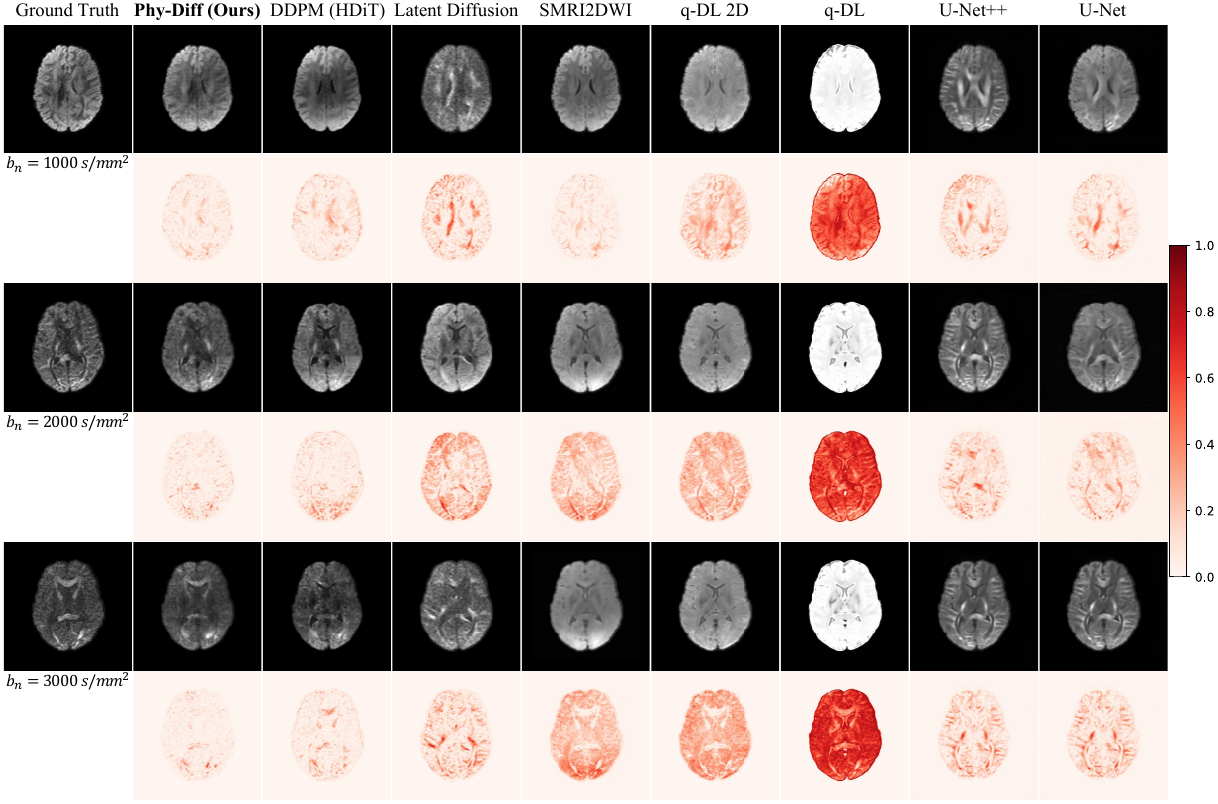}
    \caption{Examples of the generated dMRI. In each b-value block, the first row contains the ground truth and synthesized images and the second row is the error maps.}
    \label{fig:enter-label}
\end{figure}

\subsubsection{Quantitative Results.} We generated dMRI based on three different b-values, and their combined set (arbitrary $b_n$). The results showed that for any b-value, Phy-Diff outperforms other models, with SSIM up to 52.15\% and PSNR exceeding 11.22$\;dB$, which is the same as in the situation when $b_n=1000\;s/mm^2$; In the cases of $b_n=2000\;s/mm^2$ and $b_n=3000\;s/mm^2$, SSIM also surpasses other models. However, PSNR is slightly inferior. Those high PSNR images often appear overly smooth with lower SSIM, losing important structural details. In clinical diagnosis, preserving these details is crucial.

\begin{table}[t]
  \centering
  \caption{Ablation studies on arbitrary $b_n$. \textbf{Boldface} marks the top performance.}
    \begin{tabular}{lll}
    \toprule
    &SSIM\% & PSNR \\
    \midrule
    w/o Physics-guided noise evolution & 84.24±5.86 & 24.12±2.53 \\
    w/o XTRACT Adaptation & 82.63±9.94 & 24.34±2.54 \\
    w/o Query-based conditional mapping & 75.31±15.4 & 22.84±2.46 \\
    \textbf{Phy-Diff (Ours)} & \textbf{84.27±6.17} & \textbf{24.46±2.12} \\
    \bottomrule
    \end{tabular}

  \label{ablation}
\end{table}

\subsubsection{Qualitative Results.} Each various $b_n$ block in Fig.~\ref{fig:enter-label} contains ground truth as well as synthesis results illustrated in the first row, and the error maps depicted in the second row. From the generated images, Phy-Diff shows the most comparable results to ground truth, with minimal errors shown in the heat maps, especially when $b_n$ increases. Notably, Phy-Diff excels in reproducing white tracts, showcasing superior preservation of structural integrity.

\subsection{Ablation Experiments}
We evaluate the effectiveness of each component on the Phy-Diff by individually removing it. Our results (Table~\ref{ablation}) show that each component contributes to the performance improvement, with query-based conditional mapping bringing a PSNR increase of $1.62\;dB$ and SSIM of 8.96\%.

\section{Conclusion}
In this paper, we present Phy-Diff, a DM-based dMRI generation model, with physics-guided noise evolution, query-based category condition mapping, and XTRACT condition adapter. Our experiments demonstrate that Phy-Diff achieves optimal performance under multiple b-values. Moreover, compared with other state-of-the-art models, Phy-Diff demonstrates more robust performance when accepting inputs with diverse b-vectors, b-values, and slices. Phy-Diff could serve as a tool for dMRI generation, offering the potential to reduce acquisition time while maintaining high-quality results.

\begin{credits}
\subsubsection{\discintname}
The authors have no competing interests to declare that are relevant to the content of this article.
\end{credits}

%
%
\bibliographystyle{splncs04}
\bibliography{references}
%




\end{document}